\documentclass[prd,aps,floats,twocolumn]{revtex4}
\usepackage{amssymb}
\usepackage{amsmath}
\usepackage{mathrsfs}
\usepackage{latexsym}
\usepackage[dvips]{graphicx}
\begin{document}

\title{A Spinor Theory of Gravity and the Cosmological Framework}

\author{M. Novello}
\affiliation{
Institute of Cosmology, Relativity and Astrophysics ICRA/CBPF \\
Rua Dr. Xavier Sigaud 150, Urca 22290-180 Rio de Janeiro,
RJ-Brazil}
\date{\today}
\begin{abstract}

Recently we have presented a new formulation of the theory of
gravity based on an implementation of the Einstein Equivalence
Principle distinct from General Relativity. The kinetic part of the
theory - that describes how matter is affected by the modified
geometry due to the gravitational field - is the same as in General
Relativity. However, we do not consider the metric as an independent
field. Instead, it is an effective one, constructed in terms of two
fundamental spinor fields $\Psi$ and $\Upsilon$ and thus the metric
does not have a dynamics of its own, but inherits its evolution
through its relation with the fundamental spinors. In the first
paper it was shown that the metric that describes the gravitational
field generated by a compact static and spherically symmetric
configuration is very similar to the Schwarzschild metric. In the
present paper we describe the cosmological framework in the realm of
the Spinor Theory of Gravity.


\end{abstract}

\vskip2pc
 \maketitle

\section{Introduction}
There is no doubt that the activity in the field of experimental
gravitation has increased largely in the last decades. New space
measurements and astronomical discoveries, including those of
cosmological origin are mainly responsible for this. At the basis of
any theory of gravity compatible with such observations, one has the
Einstein Equivalence Principle (EEP) which can be described
\cite{will2006} as three conditions:

a. The weak equivalence principle is valid (that is, all bodies
fall precisely the same way in a gravitational field);

b. The outcome of any local non-gravitational experiment is
independent of the velocity of the freely-falling reference frame
in which it is performed;

c. The outcome of any local non-gravitational experiment is
independent of where and when in the universe it is performed;

From the validity of this EEP one infers that  "the gravitation must
be a curved space-time phenomenon". This was implemented by Einstein
by assuming that the curvature of the space-time is related to the
stress-energy-momentum tensor of matter in space-time and by
postulating a specific form for such an equation. Taken together,
the EEP and Einstein's equation constitute the basis of a successful
program of a theory of gravity.

Is this the unique way to deal with the universality of
gravitational processes? Is the only way to implement the EEP?
Recently \cite{stg1} we proposed a new look into this old question
by arguing that it is possible to treat the metric of space-time -
that in General Relativity (GR) describes the gravitational
interaction - as an effective geometry, that is, the metric acting
on matter is not an independent field and as such does not posses
its own dynamics. Instead, it inherits one from the dynamics of two
fundamental spinor fields $\Psi$ and $\Upsilon$ which are
responsible for the gravitational interaction and from which an
effective geometry appears.

The nonlinear character of gravity should be present already at the
most basic level of these fundamental structures. It seems natural
to describe this nonlinearity in terms of the invariants constructed
with the spinor fields. The simplest way to build a concrete model
is to use the standard form of a contraction of the currents of
these fields, e.g. $J_{\mu} \, J^{\mu},$ to construct the Lagrangian
of the theory. We assume that these two fields (which are
half-integer representations of the Poincar\'e group) interact
universally with all other forms of matter and energy. As a
consequence, this process can be viewed as nothing but a change of
the metric of the space-time. In other words, the influence of these
spinor fields on matter/energy is completely equivalent to a
modification of the background geometry into an effective Riemannian
geometry $g_{\mu\nu}.$ In this aspect this theory agrees with the
idea of General Relativity theory which states that the Equivalence
Principle implies a change on the geometry of space-time as a
consequence of the gravitational interaction. However, the
similarities between the Spinor Theory of Gravity and General
Relativity stop here.

To summarize, let us stress the main steps of this program.

a. There exist two fundamental spinor fields -- which we will name
$\Psi$ and $\Upsilon;$

b. The interaction of $\Psi$ and $\Upsilon$ is described by Fermi
Lagrangian;

c. The fields  $\Psi$ and $\Upsilon$ interact universally with all
forms of matter and energy;

d. As a consequence of this coupling with matter, the universal
interaction produces an effective metric;

e. The dynamics of the effective metric is already contained in
the dynamics of $\Psi$ and $\Upsilon:$ the metric does not have a
dynamics of its own, but inherits its evolution through its
relation with the fundamental spinors.

In \cite{stg1} we presented a particular example of the effective
metric in the case of a compact spherically static object, like a
star and have shown that it is astonishingly similar to the
Schwarzschild solution of GR.

Before entering the analysis of these questions let us briefly
comment our motivation. As we shall see, the present proposal and
the theory of General Relativity have a common underlying idea: the
characterization of gravitational forces as nothing but the effect
on matter and energy of a modification of the geometry of
space-time. This major property of General Relativity remains
unchanged. The main difference concerns the dynamics that this
geometry obeys. In GR the dynamics of the gravitational field
depends on the curvature invariants; in the Spinor Theory  of
Gravity such a specific dynamics simply does not exist: the geometry
evolves in space-time according to the dynamics of the spinors
$\Psi$ and $\Upsilon.$ The metric is not a field of its own, it does
not have an independent reality but is just a consequence of the
universal coupling of matter with the fundamental spinors. The
motivation of walking down only half of Einstein's path to General
Relativity is to avoid certain known problems that still plague this
theory, including its difficult passage to the quantum world and the
questions put into evidence by astrophysics involving many
discoveries such as the acceleration of the universe, the problems
requiring dark matter and dark energy. It seems worthwhile to quote
\cite{detf} here : "Dark energy appears to be the dominant component
of the physical Universe, yet there is no persuasive theoretical
explanation for its existence or magnitude. The acceleration of the
Universe is, along with dark matter, the observed phenomenon that
most directly demonstrates that our theories of fundamental
particles and gravity are either incorrect or incomplete. Recent
observations in Cosmology are responsible for an unexpected
attitude: to take seriously the possibility of alterating Einstein'
s theory of gravity". The Spinorial Theory of Gravity presents the
possibility of a way out of these difficulties. The reason, which
will be explained later on, can be understood from the fact that in
the STG there is no direct relationship between the acceleration of
the scale factor of the universe and the matter/energy distribution,
contrary to the case of GR, in which the Friedman equation that
controls the dynamics of the universe relates the matter-energy
content to the geometry through the evolution of the scale factor
$a(t):$
$$ \frac{\ddot{a}}{a} = - \, \frac{1}{6} \, ( \rho + 3 p).$$
It follows from this equation that if the universe is accelerating,
then something very unusual must occur, like, for instance, a very
negative pressure term dominating the evolution. As we shall see,
nothing similar happens in STG, since the way in which matter
influences the dynamics of the geometry does not take such form.

In the first subsection we present the mathematical background used
in the paper and in particular the very important Pauli-Kofinki
identity. These relations allow us to obtain a set of products of
currents which will be very useful to simplify our calculations. In
section II we recall the definition of the effective metric and some
of its properties and compare with the field theory formulation of
General Relativity. In section III we present the dynamics,
separated in two parts: i) the kinetic part, which tells us how
particles move in a given gravitational field; and ii) the influence
of matter on the formation of the gravitational field. We shall see
that in what concerns  the first part, the Spinor Theory of Gravity
is completely identical to General Relativity. They differ in the
second part, once in STG there is no independent dynamics for the
geometry.
 In the field theory formulation of
 General Relativity as it was described in the fifties by Gupta,
Feynman \cite{Feynman} and others,  and more recently in \cite{GPP},
the gravitational field  can be described alternatively either as
the metric of space-time -- as in Einstein's original version -- or
as a field $\varphi_{\mu\nu}$ in an arbitrary unobservable
background geometry, which is chosen to be Minkowski (see also
\cite{GPPdesitter}). We shall see that by universally coupling the
spinor fields to all forms of matter and energy, a metric structure
appears, in a similar way to the field theoretical description of
GR. The main distinction between these two approaches concerns the
status of this metric. In General Relativity it has a dynamics
provided by a Lagrangian constructed in terms of the curvature
invariants. In our proposal, this is not the case. The metric is an
effective way to describe gravity and it appears because of the
universal form of the coupling of matter/energy of any form and the
fundamental spinors. Section IV deals with the induced dynamics. In
section V we start the new cosmological program. Section VI contains
the conclusions and comments.

\subsection{Definitions}
\protect\label{algebraic}


In this paper we deal with two spinor fields $\Psi$ and $\Upsilon.$
We use capital symbols to represent the vector and axial currents
constructed with $\Psi$ as above and lower case to represent the
corresponding terms of the  spinor $\Upsilon,$ namely,
\[
J^{\mu}\equiv \overline{\Psi} \gamma^{\mu}  \Psi
\]
\[
I^{\mu}\equiv \overline{\Psi} \gamma^{\mu} \gamma^{5} \Psi.
\]
\[
j^{\mu}\equiv \overline{\Upsilon} \gamma^{\mu}  \Upsilon
\]
\[
i^{\mu}\equiv \overline{\Upsilon} \gamma^{\mu} \gamma^{5}
\Upsilon.
\]

We use the standard convention and definitions (cf.  \cite{Elbaz}).
For completeness we recall:
\[
\bar{\Psi} \equiv  \Psi^{+} \gamma^{0}.
\]
The $\gamma^{5}$ is Hermitian and the others $\gamma_{\mu}$ obey the
Hermiticity relation
\[
\gamma_{\mu}^{+} = \gamma^{0} \gamma_{\mu} \gamma^{0}.
\]

The properties needed to analyze non-linear spinors are contained in
the Pauli-Kofink (PK) relation. These are identities that establish
a set of relations concerning elements of the four-dimensional
Clifford algebra. The main property states that, for any element $Q$
of this algebra, the PK relation ensures the validity of the
identity:
\begin{equation}
(\bar{\Psi} Q \gamma_{\lambda} \Psi) \gamma^{\lambda} \Psi  =
(\bar{\Psi} Q \Psi)  \Psi  -  (\bar{\Psi} Q \gamma_{5} \Psi)
\gamma_{5} \Psi. \protect\label{H5}
\end{equation}
for $Q$ equal to $\mathbb{I}$ , $\gamma^{\mu}$, $\gamma_{5}$ and
$\gamma^{\mu} \gamma_{5},$ respectively, where $\mathbb{I}$ is the
identity of the Clifford algebra. As a consequence of this relation
we obtain two extremely important facts:
\begin{itemize}
 \item{The norm of the currents $J_{\mu}$ and $I_{\mu}$ have the same
value and opposite sign.}
 \item{Vectors  $J_{\mu}$ and $I_{\mu}$ are orthogonal.}
\end{itemize}

This formula implies some identities which will be used later on to
simplify our calculations:
\begin{eqnarray}
J_{\mu} \, \gamma^{\mu} \, \Psi &\equiv&  ( A  + i B \gamma^{5}) \,
\Psi \nonumber \\
I_{\mu} \, \gamma^{\mu} \, \gamma^{5} \,  \Psi &\equiv& -  ( A  + i
B \gamma^{5}) \, \Psi \nonumber \\
I_{\mu} \, \gamma^{\mu} \, \Psi &\equiv&  ( A  + i B \gamma^{5}) \,
\gamma^{5}  \Psi  \nonumber \\
J_{\mu} \, \gamma^{\mu} \, \gamma^{5} \, \Psi &\equiv& - ( A  + i B
\gamma^{5}) \, \gamma^{5} \Psi, \label{27dez915}
\end{eqnarray}
where $A \equiv  \bar{\Psi} \, \Psi$ and   $B \equiv i \bar{\Psi} \,
\gamma^{5} \Psi.$

\section{The effective metric}

In \cite{stg1} we showed how to treat Gravity as the universal
interaction of two fundamental spinors $\Psi$ and $\Upsilon$  with
matter. This led to the identification of gravity as a geometric
phenomenon, at least for the kinematic part, that is, when dealing
with the question of how gravity influences matter. This form of
interaction leads to the introduction of a riemannian geometry in an
analogous way as it is done in the field theoretical description of
General Relativity, namely, in terms of the metric of the background
$\eta_{\mu\nu}$ and a symmetric second order tensor
$\varphi_{\mu\nu}:$
\begin{equation}
g_{\mu\nu} = \eta_{\mu\nu} + \varphi_{\mu\nu} \label{18agosto940}
\end{equation}
The field $\varphi_{\mu\nu}$ is chosen to be non-dimensional, that
is, we set:
\begin{equation}
\varphi_{\mu\nu} = - \frac{g_{F} \, g_{m}}{4} \,
\frac{1}{\sqrt{X}} \, (c_{\mu\nu} + c_{\nu\mu})
\label{18agosto945}
\end{equation}
where
\begin{equation*}
c_{\mu\nu} = \Sigma_{\mu} \, \Pi_{\nu}
\end{equation*}
and the vectors  $\Sigma_{\mu},  \, \Pi_{\mu}$ are constructed in
terms of the currents of the spinors, through the definitions
$$\Sigma_{\mu} \equiv J_{\mu} + j_{\mu} +  I_{\mu} + i_{\mu}$$
and
$$\Pi_{\mu} \equiv  J_{\mu} + j_{\mu} + \beta \, (I_{\mu} +
i_{\mu}).$$ $ X $ is defined as the trace \cite{comentario}
$$X
\equiv c^{\mu\nu} \eta_{\mu\nu} =
 \Sigma^{\mu}\, \Pi_{\mu}.$$  Note that Minkowski geometry is the
unique metric that we deal with in what concerns $\Psi$ and
$\Upsilon.$  The effective metric $g_{\mu\nu}$ appears only when
considering the evolution of matter. The asymmetry exhibited by
vectors $\Sigma_{\mu}$ and $\Pi_{\mu}$ is a consequence of the way
we constructed these terms: from the association of a
self-interacting Heisenberg potential with a modification of the
internal connection, as demonstrated in the previous paper. This is
the only free parameter of the theory. We shall see in the next
section that the case in which  $\beta = 1$  a number of interesting
special properties appear. To quote some: it was shown in
\cite{stg1} that the gravitational field of a static spherically
symmetric configuration coincides with the Schwarszchild solution of
General Relativity; the Heisenberg self-interaction disappears; the
coupling between the fundamental spinor fields coincides with low
energy weak interaction.
 The constant $g_{F}$ has the
 same dimensionality as Fermi constant and $ g_{m} $ has the
dimensionality of $(energy)^{-1}.$ We note that in the previous
article the constant $g_{m}$ was named $\lambda.$

 Let us emphasize that this form of the
construction of the metric tensor is an exact one: it is not an
approximation. This implies that the inverse contra-variant
expression $g^{\mu\nu}$ defined by $g^{\mu\alpha} \, g_{\alpha\nu} =
\delta^{\mu}_{\nu}$ is an infinite series:
\begin{equation}
g^{\mu\nu} = \eta^{\mu\nu} - \varphi^{\mu\nu}+ \varphi^{\mu\alpha}
\, \varphi_{\alpha} {}^{\nu} + ... \label{6dez910}
\end{equation}

 In order to be clearly
understood, let us pause for a while to very briefly describe the
so-called field theoretical description of General Relativity while
 emphasizing that such a description is not related to our proposal.

\subsection{The universal coupling: gravity}

The field theoretical approach of GR goes back to the fact that
Einstein dynamics of the curvature of the Riemannian metric of
space-time can be obtained as a sort of iterative process, starting
from a linear theory of a symmetric second order tensor
$\varphi_{\mu\nu}$ and by an infinite sequence of self-interacting
processes leading to a geometrical description. The definition of
the metric is the same as above. Other definitions were also used
(see for instance \cite{GPP} for an analysis of the convenience of
these alternative non-equivalent definitions). Although these
theories can be named "field theories", they contain the same metric
content of General Relativity,  disguised in a non geometrical form.
The framework of the Spinor Theory of Gravity is totally different.
Let us emphasize that we are not presenting a dynamics for the
metric in the sense of such field theories. Instead, the geometry is
understood as an effective one, in the sense that it is the way
gravity appears for all forms of matter and energy. However, its
evolution is provided by the fundamental spinor fields $\Psi$ and
$\Upsilon$. These field theories of gravitation teach us how to
couple the tensor field $\varphi_{\mu\nu}$ with matter terms in
order to guarantee that the net effect of this interaction produces
the desired modification of the metric structure. This idea will
guide us when coupling the two fundamental spinors with all forms of
matter and energy in order to obtain the same equivalent
interpretation of the identification of the gravitational field with
the metric of the space-time.

\section{Dynamics}

The coupling of matter to gravity is provided by the identification
of the gravitational field with the geometry. This means that we
have to modify the matter Lagrangian in the Minkowski background by
changing $\eta_{\mu\nu}$ to $g_{\mu\nu} \equiv \eta_{\mu\nu} +
\varphi_{\mu\nu}.$ This part of the action -- which answers the
question of how gravity acts on matter -- has the same structure as
in General Relativity. However, in STG the geometry is an effective
one and does not have a dynamics of its own. In order to obtain the
evolution of the metric we have to look into the dynamics of the
spinors.

\subsection{The behavior of matter in a given gravitational field}
Assume that in the Special Theory of Relativity the dynamics of a
certain matter distribution is given by a Lagrangian $L_{m}.$ The
field theoretical description of General Relativity, describes its
interaction with gravity using the Equivalence Principle, also known
as the minimal coupling principle. This means substituting all terms
in the action $S_{0}$ in which the Minkowski metric \cite{sabado}
$\gamma_{\mu\nu}$ appears  by the effective metric $g_{\mu\nu}$ and
its inverse $g^{\mu\nu}.$ As an example consider a scalar field
$\Phi.$ In the Minkowski background,  its dynamics is provided by
\begin{equation*}
S_{0} = \int \sqrt{-\gamma} \, B^{\mu\nu} \, \gamma_{\mu\nu} =
\int \sqrt{-\gamma} \, \partial^{\mu}\Phi \,
\partial^{\nu}\Phi \, \gamma_{\mu\nu},
\end{equation*}
where $\gamma \equiv \det \gamma_{\mu\nu}.$  In this case
$B^{\mu\nu}$ can be written in terms of the energy-momentum tensor
defined as
\begin{equation*}
E^{\mu\nu} = \frac{2}{\sqrt{-\gamma}} \, \frac{\delta}{\delta
\gamma_{\mu\nu}} \left( \sqrt{-\gamma} L \right).
\end{equation*}
Indeed, a direct calculation yields
\begin{equation*}
E^{\mu\nu} = \partial_{\alpha}\Phi \partial_{\beta}\Phi \,
\gamma^{\alpha\mu} \gamma^{\beta\nu} - \frac{1}{2}
\partial_{\lambda}\Phi \partial_{\sigma}\Phi \,
\gamma^{\lambda\sigma} \gamma^{\mu\nu}
\end{equation*}
immediately implying
\begin{equation*}
B^{\mu\nu} = E^{\mu\nu} - \frac{1}{2} E \gamma^{\mu\nu},
\end{equation*}
where $ E \equiv E^{\mu\nu} \, \gamma_{\mu\nu}.$
 The corresponding action, including the gravitational
 interaction, is obtained by replacing $\gamma_{\mu\nu}$ and its inverse
$\gamma^{\mu\nu}$ with the corresponding $g_{\mu\nu} =
\gamma_{\mu\nu} + \varphi_{\mu\nu}$ which yields
\begin{equation*}
S = \int \sqrt{- \gamma} \, \omega \, \partial^{\mu}\Phi \,
\partial^{\nu}\Phi \, g_{\mu\nu},
\end{equation*}
where we have used the same definition as in the field theory of
gravity, namely,  $ \omega \equiv \sqrt{-g}/\sqrt{-\gamma},$ and
$g = det \, g_{\mu\nu}.$ In this case
\begin{equation*}
B^{\mu\nu} = \omega \, [E^{\mu\nu} - \frac{1}{2} E g^{\mu\nu}].
\end{equation*}
Thus, for any kind of matter interacting with the gravitational
field,  the action is provided by the golden rule of GR, namely
\begin{equation}
S = \int \sqrt{-\gamma} \, \omega \, L_{M} = \int \sqrt{-g} \, L_{M}
\label{30nov}
\end{equation}
where the corresponding energy-momentum tensor is given by
\begin{equation*}
T^{\mu\nu} = \frac{2}{\sqrt{-g}} \, \frac{\delta}{\delta
g_{\mu\nu}} \left( \sqrt{-g} L \right). \label{30nov2}
\end{equation*}
It follows \cite{landau} that this quantity is divergence-free, in
the effective metric $g_{\mu\nu},$ that is, $T^{\mu\nu}{}_{; \nu} =
0.$

This way of coupling matter with the fundamental spinors guarantees
that, kinematically, the behavior of any kind of matter (and energy)
in the Spinor Theory of Gravity is the same as in General
Relativity: free particles follow geodesics in a prescribed
geometry, as the manifestation of gravitational interaction.

Let us now turn to the influence of matter on the gravitational
field.  The dynamics of the gravitational field is completely
distinct in these two theories. In General Relativity, the metric
obeys a dynamics generated by the Hilbert-Einstein Lagrangian
$$ S_{HE} = \frac{1}{k_{e}} \int \,  \sqrt{-g} \, R d^{4}x.$$
Nothing similar occurs in the Spinorial Theory of Gravity. The
metric does not have a specific dynamics, but instead obeys the
evolution dictated by its relationship with the dynamics of the
fundamental spinors.

\section{Generating the gravitational field}
%
The dynamics presented in \cite{stg1} contains the following terms:
\begin{equation}
L = L (\Psi) + L(\Upsilon) + L_{int} ( \Psi, \Upsilon) + L_{mat}.
\label{27dez10}
\end{equation}
 We concentrate our analysis on the equation for the spinor $\Psi.$
 The corresponding equation for the other field $\Upsilon$ is obtained similarly by
 substituting $\Psi$ by $\Upsilon.$ We have:
\begin{equation}
i  \gamma^{\mu} \partial_{\mu} \Psi + g_{F} \, \gamma_{\mu} \left(
C^{\mu} + D^{\mu} \gamma^{5} \right) \Psi =0 \label{29nov2006}
\end{equation}
We write in the equivalent compact form:
\begin{equation}
i  \gamma^{\mu} \partial_{\mu} \Psi + g_{F} \,  {\cal{H}} \, \Psi =0
\label{27dez1010}
\end{equation}
We separate the interaction in three parts:
\begin{equation}
  {\cal{H}}= {\cal{H}}_{s} +  {\cal{H}}_{o} +  {\cal{H}}_{m},
  \label{27dez14}
  \end{equation}
  which represents the self-interaction ${\cal{H}}_{s},$  the interaction
with the other spinor  ${\cal{H}}_{o}$ and the influence of matter
${\cal{H}}_{m}.$ Thus, the quantities $C^{\mu}$ and $D^{\mu}$ are
separated in three parts, according to their origin in the process
of interaction. Let us summarize what was pointed out in the
previous paper \cite{stg1}.

\subsection{Self-interaction}
 We have:
\begin{equation}
{\cal{H}}_{s} \, \Psi = ( 1 - \beta ) ( A + i B \, \gamma^{5} ) \,
\Psi
\end{equation}
which implies
\begin{eqnarray}
C^{\mu}_{s} &\equiv& J^{\mu} + \frac{1 + \beta}{2} \, I^{\mu}
\nonumber \\
D^{\mu}_{s} &\equiv& \frac{1 + \beta}{2} \, J^{\mu} + \beta I^{\mu}
\end{eqnarray}
This term, which contains only quantities constructed with the
spinor $\Psi$ itself, is given by the quartic Heisenberg Lagrangian,
the simplest non-linear covariant term which can be constructed with
a spinor field. The Lagrangian is
\begin{equation}
L_{s} = \frac{i}{2} \bar{\Psi} \gamma^{\mu}
\partial_{\mu} \Psi - \frac{i}{2} \partial_{\mu} \bar{\Psi}
\gamma^{\mu} \Psi - V(\Psi). \protect\label{H1}
\end{equation}
Potential $V$ is constructed with the two scalars that can be formed
with $\Psi$, which are $A$ and $B.$ Thus the  Heisenberg potential
 is
\begin{equation}
V = \frac{1- \beta}{2} \, g_{F} \, \left( A^{2} + B^{2} \right).
 \protect\label{H3}
\end{equation}
 Note that Pauli-Kofink identity implies that
$$A^{2} + B^{2} = J_{\mu} J^{\mu}.$$
For the case $\beta = 1,$ the self-interacting Heisenberg term
vanishes.

\subsection{Interaction with the other fundamental spinor
$\Upsilon$ } We have:
\begin{eqnarray}
{\cal{H}}_{o} \, \Psi &=& \gamma_{\mu}  \left( j^{\mu} + \frac{( 1
+ \beta)}{2} \, i^{\mu}  \right) \, \Psi \nonumber \\
&+&   \gamma_{\mu} \gamma^{5} \left( \frac{( 1 + \beta)}{2} \,
j^{\mu} + \beta \, i^{\mu}  \right) \, \Psi
\end{eqnarray}

The interacting Lagrangian is provided by
\begin{eqnarray}
L_{o} &=& g_{F} \, \{  J_{\mu} j^{\mu} + \beta I^{\mu}i_{\mu} \}
\nonumber \\
&+& \frac{g_{F}}{2} \, (1 + \beta) (J^{\mu}i_{\mu} + I^{\mu}
j_{\mu} ). \label{18agosto701}
\end{eqnarray}
In the case $\beta = 1$ the interaction assumes the reduced form
\begin{equation*}
L_{F} =  g_{F} \, \overline{\Psi} \gamma^{\mu} ( 1 + \gamma^{5} )
\Psi
 \, \, \, \overline{\Upsilon} \gamma_{\mu} (1 + \gamma^{5} )
 \Upsilon. 
\end{equation*}
 This term is similar to the Lagrangian of weak interaction
processes in the ancient Fermi treatment. It appears here as the
natural covariant interaction between the two fundamental spinors.
The Fermi constant $g_{F}$ appears for dimensionality reasons. The
presence of such constant in the realm of gravitational world may
seem very unusual . However, an interesting remark attributed to W.
Pauli \cite{jauch} makes this identification less strange. It is
generally argued that, as far as gravity is concerned, the quantity
$10^{-33}cm$ is an important one. This number appears very naturally
by simple dimensional analysis and in certain scientific communities
this length is associated to the appearance of quantum gravitational
processes. Its expression contains three ingredients: the
relativistic quantity c (the light velocity), the Heisenberg
constant $\hbar$ and a typical gravity representative provided by
Newton 's constant $g_{N},$ yielding the Planck-Newton constant:
$$ L_{PN} = \sqrt{ \frac{h \, g_{N}}{c^{3}}}.$$
A similar quantity cannot be constructed with the other known long
range field (electrodynamics), but it can be defined for the weak
interaction. In this case we have only to exchange $g_{N}$
 by the Fermi constant, yielding the definition of what we
 call the Planck-Fermi length:
$$ L_{PF} = \sqrt{ \frac{g_{F}}{\hbar c}}.$$
Now Pauli remarks that this quantity is equal to $10^{-16}cm,$ the
square-root of the Planck-Newton value \cite{pauli1}. It is clear
that such a coincidence depends on the units used. The original
argument, which in a sense was re-taken by Dicke in 1957 deals with
the so-called " natural system of units" for the high energy physics
community, that is for  $\hbar = c = 1$ and by taking a specific
unit of mass (the electron mass in the Dicke's choice).

In the case of the interaction of the fundamental spinors, the
vectors $C^{\mu}, D^{\mu}$ are given by
\begin{eqnarray}
C^{\mu}_{o} &\equiv& j^{\mu} + \frac{1 + \beta}{2} \, i^{\mu}
\nonumber \\
D^{\mu}_{o} &\equiv& \frac{1 + \beta}{2} \, j^{\mu} + \beta i^{\mu}
\end{eqnarray}

\subsection{The effect of matter in the generation of gravity}

This term is provided by (\ref{30nov}) inspired by the Equivalence
Principle that states that the matter interacts only through the
effective metric $g_{\mu\nu}.$ Variation of the spinor $\Psi$ in
equation (\ref{30nov})  yields
\begin{eqnarray}
\delta S &=& - \frac{1}{2} \int \sqrt{-g} \, T^{\mu\nu} \,
\delta g_{\mu\nu} \nonumber \\
&=& - \frac{1}{2} \int \sqrt{-g} \, T^{\mu\nu} \,
\delta \varphi_{\mu\nu} \nonumber \\
&=& \frac{g_{F}\, g_{m}}{2} \, \int \sqrt{-g} \, T^{\mu\nu} \,
\delta \left( \frac{\Sigma_{\mu} \Pi_{\nu}}{\sqrt{X}} \right)
\end{eqnarray}
where we used (\ref{18agosto940}) and (\ref{18agosto945}). Then we
can write
$$ \delta S = \frac{g_{F}\, g_{m}}{2} \, \left(I_{1} + I_{2} + I_{3} \right) $$ where
\begin{eqnarray}
 I_{1} &=&  \int \sqrt{-g} \, T^{\mu\nu} \, \Sigma_{\mu} \, \Pi_{\nu}
\, \delta \frac{1}{\sqrt{X}} \nonumber \\
I_{2} &=&  \int \sqrt{-g} \, T^{\mu\nu} \, \frac{1}{\sqrt{X}} \,
\Sigma_{\mu} \delta \Pi_{\nu}  \nonumber \\
I_{3} &=&  \int \sqrt{-g} \, T^{\mu\nu} \, \frac{1}{\sqrt{X}} \,
\Pi_{\nu} \delta \Sigma_{\mu}
\end{eqnarray}
A direct calculation yields:
\begin{eqnarray}
C^{\mu}_{m} &\equiv&   \,  \frac{g_{m}}{4} \omega \, \left(
 - \, \frac{1}{2} \, \Phi \, \xi^{\mu} +  E^{\mu} + H^{\mu} \right) \nonumber \\
D^{\mu}_{m} &\equiv&   \,  \frac{g_{m}}{4} \omega \, \left( - \,
\frac{1}{2} \,  \Phi \, \eta^{\mu} + \beta E^{\mu} + H^{\mu} \right)
\label{12dez545}
\end{eqnarray}
where
$$ E^{\mu} = \frac{1}{\sqrt{X}} \, T^{\mu\nu} \Sigma_{\nu}, $$
$$ H^{\mu} = \frac{1}{\sqrt{X}} \, T^{\mu\nu} \Pi_{\nu}, $$
$$ \Phi = \frac{1}{X^{3/2}} \, T^{\mu\nu} \Sigma_{\mu} \Pi_{\nu}. $$
$$ \xi^{\mu} = \Pi^{\mu} + \Sigma^{\mu} , $$
$$\eta^{\mu} = \Pi^{\mu} + \beta \, \Sigma^{\mu}.$$

In the STG this is how matter generates gravitational fields.

The most important task now is to analyze the consequences of this
theory. In a previous paper \cite{stg1} we started by studying the
effective metric generated by the gravitational process in the
neighborhood of a massive object, such as a star. Here we
concentrate on cosmology.

For later use it is useful to separate this matter influence into
three parts using the notation of equation (\ref{27dez14}):
\begin{equation}
{\cal{H}}_{m} = {\cal{T}}_{s} + {\cal{T}}_{o} + {\cal{T}}_{m}
\label{27dez1410}
\end{equation}
where
\begin{eqnarray}
{\cal{T}}_{s} &=& - \frac{g_{m}}{2} \, \omega \, \Phi \, ( 1 - \beta
)
\, ( A + i B \, \gamma^{5} ) \nonumber \\
&=&  - \frac{g_{m} \omega \Phi}{2} \, {\cal{H}}_{s}
\label{27dez1411}
\end{eqnarray}

\begin{eqnarray}
{\cal{T}}_{o} &=& - \frac{g_{m}}{2} \, \omega \, \Phi \,  j^{\mu}
\left( \gamma_{\mu} + \frac{(1 + \beta)}{2} \, \gamma_{\mu}
\gamma^{5}  \right) \nonumber \\
&-&  \frac{g_{m}}{2} \, \omega \, \Phi \, i^{\mu} \, \left( \frac{(1
+ \beta )}{2} \, \gamma_{\mu} + \beta \, \gamma_{\mu} \gamma^{5}
\right) \nonumber \\
&=& - \frac{g_{m}}{2}\, \omega \, \Phi \, {\cal{H}}_{o},
\label{27dez1412}
\end{eqnarray}

\begin{eqnarray}
{\cal{T}}_{m} &=&  \, \frac{g_{m}}{4} \, \omega \,
\gamma_{\mu} \, \left( E^{\mu} + H^{\mu} \right) \nonumber \\
&+& \, \frac{g_{m}}{4} \, \omega \, \gamma_{\mu} \gamma^{5} \,
\left( \beta \, E^{\mu} + H^{\mu}  \right). \label{27dez1413}
\end{eqnarray}
The origin of these terms is very similar to the other expression.
Indeed, $ {\cal{T}}_{s}$ is proportional to ${\cal{H}}_{s};$ the
term $ {\cal{T}}_{o}$ is proportional to  ${\cal{H}}_{o}.$ This
suggests treating the third term in such a way that it can be
reduced to a combination of both terms. We postpone this analysis to
another place. Here we concentrate on a cosmological scenario in
which matter is not important for the generation of the
gravitational field.

\section{Cosmology in the STG}
We now present a solution of the Spinor Theory of Gravity which
represents an empty spatially homogeneous and isotropic expanding
 universe. This case shows a
 net distinction between the properties of the STG and GR. Indeed, in
the case of General Relativity it is not possible to conciliate an
empty universe with a FRW type geometry. On the other hand, we shall
show that if we  neglect the influence of matter, the Spinor Theory
of Gravity allows for a cosmological solution which represents a
geometry that is spatially homogeneous and isotropic.

\subsection{The Milne expanding universe}

In order to find a solution to the STG equations of motion, the
first step is to make a convenient choice of the coordinate system
that describes the unobserved auxiliary geometry where the spinors
live.  In the case of cosmology, the structure of the effective
geometry should have the property of being non-static, spatially
homogeneous and isotropic.  In order to simplify our calculations it
is convenient to choose a coordinate system that represents
Minkowski background geometry which already exhibits such property.
This led us to deal with Milne description of the flat space-time
manifold. We set
\begin{equation}
ds^{2} = dt^{2} - t^{2} \, \left( d\chi^{2} + sinh\chi^{2}
(d\theta^{2} -  sin^{2}\theta d\varphi^{2}) \right), \label{1dez06}
\end{equation}
In consequence, the $\gamma_{\mu}$'s are given in terms of the
constant $\widetilde{\gamma}_{\mu}$ as follows:
\begin{eqnarray}
\gamma_{0} &=&  \widetilde{\gamma}_{0} \nonumber \\
\gamma_{1} &=&  t \, \widetilde{\gamma}_{1}  \nonumber \\
\gamma_{2} &=&  t \, \sinh \chi \, \widetilde{\gamma}_{2} \nonumber \\
\gamma_{3} &=&  t \, \sinh \chi \, \sin \theta \,
\widetilde{\gamma}_{3}. \nonumber
\end{eqnarray}
For later use we display our convention of the constant
$\gamma_{\mu}$'s:
\[
\widetilde{\gamma}^{0} =  \left(
\begin{array}{cccc}
I_{2} &
   0 \nonumber \\
 0 &
  -\, I_{2} &
    \nonumber \\
  \end{array}
\right) \nonumber
\]

\[
\widetilde{\gamma}_{k} =  \left(
\begin{array}{cccc}
0 &
   \sigma_{k} \\
 -\sigma_{k} &
  0 &
    \\
  \end{array}
\right) \nonumber
\]
\[
\gamma^{5} =  \left(
\begin{array}{cccc}
0 &
   I_{2}  \\
 I_{2}&
  0 &
    \\
  \end{array}
\right). \nonumber
\]

This form was obtained by using the property
\begin{equation}
\gamma_{\mu} \, \gamma_{\nu} + \gamma_{\nu} \, \gamma_{\mu} = 2
g_{\mu\nu} \, \mathbb{I}. \label{11agosto20}
 \end{equation}  Note
that, from now on, we write $1$ instead of $\mathbb{I}$ to represent
the identity of the Clifford algebra. Since we are led to use a
coordinate system that is not Euclidean, we must use the generalized
covariant derivative. In the case of the original Fock-Ivanenko
condition (i.e., vanishing of the covariant derivative of the
$\gamma_{\mu}$) one obtains:
\begin{equation}
\Gamma_{\mu}^{0} = \frac{1}{8} \, \left[ \gamma^{\alpha} \gamma_{\mu
\, ,\alpha} - \gamma_{\mu \, ,\alpha} \gamma^{\alpha} +
\Gamma^{\epsilon}_{\mu\nu} \, (\gamma_{\epsilon} \gamma^{\nu} -
\gamma^{\nu} \gamma_{\epsilon}) \right]. \label{11agosto2940}
\end{equation}
The index $(0)$ in $\Gamma_{\mu}$ is just a reminder that we are
dealing with a Minkowski background in an arbitrary system of
coordinates. We can globally annihilate such a connection by moving
to a Euclidean constant coordinate system. Using these quantities,
we obtain the unique non identically null background FI connection:
\begin{eqnarray*}
\Gamma_{1}^{(0)} &=& - \, \frac{1}{4} \, \widetilde{\gamma}_{0} \,
\widetilde{\gamma}_{1} \nonumber \\
\Gamma_{2}^{(0)} &=& - \, \frac{1}{4} \,  \sinh \chi \,
\widetilde{\gamma}_{0} \, \widetilde{\gamma}_{2} + \frac{1}{4}
cosh\chi \,
  \widetilde{\gamma}_{1} \, \widetilde{\gamma}_{2} \nonumber
  \\
\Gamma_{3}^{(0)} &=& - \, \frac{1}{4} \,  \sinh \chi \,sin\theta
\widetilde{\gamma}_{0} \, \widetilde{\gamma}_{3} + \frac{1}{4} \,
\cosh \chi \, \sin \theta \, \widetilde{\gamma}_{1} \,
\widetilde{\gamma}_{3 } \nonumber \\  &+& \frac{1}{4} \, \cos \theta
\, \widetilde{\gamma}_{2} \, \widetilde{\gamma}_{3 }
\label{13dez1155}
\end{eqnarray*}

\subsection{The cosmological metric}

To simplify our presentation we will consider the particular case
in which $\beta = 1.$  The equations of motion in this case are
\begin{eqnarray}
i \gamma^{\mu} \, \partial_{\mu} \Psi &+&  \gamma^{\mu} \,
\Gamma_{\mu}^{(0)} \Psi \nonumber \\
&+& g_{F} \, \gamma_{\mu} \left( C^{\mu} + D^{\mu} \gamma^{5}
\right) \Psi =0 \label{13agosto1510}
\end{eqnarray}
where
\begin{equation}
C^{\mu}_{o} = D^{\mu}_{o} =  j^{\mu} +  i^{\mu}
\end{equation}
and an analogous equation for $\Upsilon.$

 This is a non linear system that must be solved in order
 to obtain the effective metric. We look for a solution of the form
\begin{equation*}
\Psi =  f(t) \, e^{i R(t)} \, e^{i \,H(\chi)} \, e^{i \,L
(\theta)} \Psi^{0}
\end{equation*}
\begin{equation*}
\Upsilon = f^{'}(t) \,e^{i R^{'}(t)} \, e^{i \, H^{'}(\chi)} \,
e^{i\, L^{'} (\theta)}  \, \Upsilon^{0} \label{1dez1225}
\end{equation*}
where $\Psi^{0}$ and $\Upsilon^{0}$ are constant spinors. This
choice implies immediately that the currents constructed with
these spinors may depend only on time. Substituting the above
expressions in the fundamental equation we obtain
\begin{equation}
{\cal{W}} \, \Psi = 0
\end{equation}
where we defined
\begin{eqnarray}
{\cal{W}} & \equiv & i \widetilde{\gamma}_{0}  \, ( \frac{1}{f}
\frac{df}{dt} + i \frac{dR}{dt} ) +  \frac{1}{t} \,
\widetilde{\gamma}_{1}
\frac{dH}{d\chi}  \nonumber \\
&+&  \frac{1}{t \sinh\chi} \, \widetilde{\gamma}_{2}
\frac{dL}{d\theta}   + \frac{3}{4t}  \widetilde{\gamma}_{0}
\nonumber \\
&-& \frac{1}{2t} \, \cot(\chi) \widetilde{\gamma}_{1} -
\frac{1}{4t} \, \frac{\cot\theta}{\sinh\chi}\widetilde{\gamma}_{2}
 \nonumber \\
 &-& g_{F} \,\widetilde{\gamma}_{\mu} (1 + \gamma^{5} ) ( j_{\mu} + i_{\mu} )
\end{eqnarray}
Thus,  $R, H$ and  $L$ are given by
\begin{eqnarray}
R &=& \epsilon \, ln \sqrt{t} \nonumber \\
H &=& \frac{1}{4} \, ln sin \chi + i \alpha \chi \gamma^{0} \nonumber \\
L &=& \frac{1}{4} \, ln sin \theta.
\end{eqnarray}
which yield
\begin{equation}
\Psi = f(t) \, e^{i\varepsilon ln \sqrt{t}} \, e^{\frac{i}{2} ln
 sin\chi + i \alpha\chi \gamma^{0}} \, e^{\frac{i}{4} ln sin\theta}  \, \Psi^{0}, \label{14agosto1415}
\end{equation}
\begin{equation}
\Upsilon = f^{'}(t) \, e^{i\varepsilon^{'} ln \sqrt{t}} \,
e^{\frac{i}{2} ln
 sin\chi + i \alpha\chi \gamma^{0}} \, e^{\frac{i}{4} ln sin\theta}  \, \Upsilon^{0}
\label{1dez1230}
\end{equation}
where $\varepsilon, \varepsilon^{'}, \alpha$ and $\alpha^{'}$  are
constants. From this form it follows immediately that the dependence
on $\chi$ and $\theta$ disappear in both (vector and axial)
currents. The functions $f$ and $f^{'}$ differ by a constant factor
and both satisfy the equation

\begin{equation}
\frac{1}{f^{3}} \, \frac{df}{dt}  = \mu_{0}, \label{12dez1035}
\end{equation}
where $\mu_{0}$ is a constant. Thus, the function $f$ is
proportional to $ t^{- \, 1/2}.$

 Since we are solely interested in the behavior of the geometry, we note
 that only function $f$ is of interest. In order to find a solution
such that the form of the geometry represents a spatially
homogeneous and isotropic geometry all the components of the
currents but $\Sigma_{0}$ and $\Pi_{0}$must be zero. This is
achieved by imposing the restrictions $ J_{k} + j_{k} = 0$ and $
I_{k} + i_{k} = 0.$  The constant spinor $\Psi^{0}$ (respectively
$\Upsilon^{0}$) satisfies the equation:

\begin{equation}
i \mu_{0} \gamma^{0}\, \Psi^{0} + {\cal{H}}_{o} \Psi^{0} = 0,
\end{equation}
which is the compatibility condition that must be satisfied for the
constant spinor $\Psi^{0}.$

\subsection{The fundamental equation for the geometry}

From these expressions and using the construction
(\ref{18agosto945}) the effective metric representing the
gravitational interaction becomes
\begin{equation}
ds^{2} = ( 1 - n^{2} \, f^{2} ) dt^{2} - t^{2} d\sigma^{2}.
\label{12dez500}
\end{equation}
In order to describe such geometry in the standard gaussian global
time we define the cosmical time $T$ by
\begin{equation}
dT = \sqrt{1 - n^{2} \, f^{2}} \, dt. \label{12dez1435}
\end{equation}
In this way, the effective geometry takes the conventional FRW
form
\begin{equation}
ds^{2} = dT^{2}  - a^{2} d \sigma^{2},
\label{12dez520}\end{equation} in which the variable $t$ is now a
function of the gaussian time $T$  and is re-named by the
substitution $ a \equiv t(T).$ Using the solution of $f$ we can
write the implicit dependence of the scale factor on the global
time:
\begin{equation}
T = a \, \sqrt{1 - \frac{Q^{2}}{a}} + \frac{Q^{2}}{2}  \, log \,
{\cal{G}}
\end{equation}
where

 $${\cal{G}} \equiv \frac{\sqrt{a} - \sqrt{a - Q^{2}}} {\sqrt{a}
+ \sqrt{a - Q^{2}}} \,.$$

Before continuing let us comment on the properties of the expanding
universe associated with this geometry as oposed to the similar
result in General Relativity. In GR the equations of motion relate
the acceleration of the scale factor to the density and pressure
through the Raychaudhuri equation
$$ \frac{d^{2}a}{dT^{2}} = - \, \frac{1}{6} \, (\rho + 3p). $$
This is precisely the basis of the recent difficulties of explaining
the origin of the acceleration of the universe. Let us point out
that since the Spinor Theory of Gravity does not have a direct
equation of motion relating the metric evolution to the matter
sources this kind of difficulty does not appear. Thus one can expect
that STG may in principle conciliate the observed acceleration with
non-exotic matter. This will be treated in a future paper. We obtain
from the above equation the expansion factor or Hubble parameter
\begin{equation}
\frac{da}{dT} = \frac{1}{\sqrt{ 1 - n^{2} \, f^{2}}}
\end{equation}
and the acceleration
\begin{equation}
\frac{d^{2}a}{dT^{2}} = \left( \frac{n f^{2}}{ 1 - n^{2} \, f^{2}}
\right)^{2} \, \mu_{0}.  \label{12dez1055}
\end{equation}

Thus the sign of the acceleration factor depends only on the
constant $\mu_{0}.$ For a real function $f$ equation
(\ref{12dez1035}) implies that $\mu_{0} < 0$ and in this case the
scale factor is such that $ \frac{d^{2}a}{dT^{2}} < 0.$

\subsection{The equivalence of General Relativity}

Having obtained the explicit empty cosmological solution of the
effective geometry produced in the realm of Spinor Theory of
Gravity, one could envisage the equivalent matter-energy
distribution, in the realm of General Relativity, that produces the
same geometry. For this we must write the equations of motion of
this metric in GR. They are given by a fluid with density of energy
and pressure that we named, respectively, $\rho_{E}$ and $p_{E}.$ By
definition, they satisfy Einstein's equations:
\begin{eqnarray}
\rho_{E} &=& 3 \left(\frac{\dot{a}}{a}\right)^{2} -
\frac{3}{a^{2}}
\nonumber \\
p_{E} &=&  - 2 \frac{\ddot{a}}{a} -
\left(\frac{\dot{a}}{a}\right)^{2} + \frac{1}{a^{2}}
\end{eqnarray}
Using the above expressions for $a(T)$ we obtain for the expansion
and the acceleration:

$$ \frac{da}{dt} = \left( 1 - \frac{Q^{2}}{a} \right)^{- \,1/2},$$
and
$$ \frac{d^{2}a}{dt^{2}} = - \, \frac{Q^{2}}{a^{2}} \, \left( 1- \frac{Q^{2}}{a} \right)^{- \,2}.$$
Then, the equivalent density of energy and pressure in GR
are
\begin{eqnarray}
\rho_{E} &=& 3 \, Q^{2} \, a^{- \, 2} \, \left( a - Q^{2} \right)^{- 1} \nonumber \\
p_{E} &=&  Q^{4} \, a^{- \, 2} \, \left( a - Q^{2} \right)^{- 2}.
\end{eqnarray}
that is, one obtains an equatiion of state given by
$$ p_{E} = \frac{1}{9} \, \left( \rho \, a \right) ^{2}.$$

We remark that such an equation of state is not only non-linear, but
it also depends on the scale factor, changing continuously as the
universe expands. At very late times, when the scale factor is big
enough that one can neglect the constant $Q^{2}$ we obtain the
approximated expressions:
\begin{eqnarray}
\rho_{E} &\approx& a^{- 3}  \nonumber \\
p_{E} &\approx& a^{- 4}.
\end{eqnarray}
Let us just point out that this solution, like the standard FRW
geometry, contains a particle horizon. Indeed, setting the
definition:
\begin{equation}
\mathfrak{I} \equiv \int \frac{dT}{a}
\end{equation}
it follows
\begin{equation}
\mathfrak{I} = -  \, log \left( \frac{1 - cos y}{1 + cos y}\right)
- 2 cos y
\end{equation}
for $y \equiv arcsin \, Q/ \sqrt{a}.$ Thus the integral $
\mathfrak{I}$ converges in the domain $ Q^{2} < a < a_{0},$ for
finite $a_{0},$ showing that this geometry has a particle horizon.

\section{Conclusion}
In the present paper we have examined some properties of a new
formalism to describe gravity. We have shown that there is an
alternative way to implement the Equivalence Principle in which the
geometry acting on matter is not an independent field, and as such
does not posses its own dynamics. Instead, it inherits one from the
dynamics of two fundamental spinor fields $\Psi$ and $\Upsilon$
which are responsible for the gravitational interaction and through
which the effective geometry appears. We have presented a specific
model by using Heisenberg equation of motion for the
self-interacting spinors. This equation of motion can be understood
in terms of a modification of the internal connection as seen by
$\Psi$ and $\Upsilon$ and only by these two spinors, as it was
described in \cite{stg1}. This dynamics, which involves not only the
self terms but also a specific coupling among these two fields,
provides an evolution for the effective metric (constructed in terms
of these spinors) which is the way the fields $\Psi$ and $\Upsilon$
interact with all other forms of matter and energy. In a previous
work we found a solution for the effective metric in the case of a
static spherically symmetric configuration. The result is similar as
in General Relativity, showing the existence of horizon and the
possibility of existence of Black Hole. In the present paper we have
turned our analysis to cosmology. We found a solution of the
fundamental spinor fields that corresponds to the interaction of
$\Psi$ to $\Upsilon$ without any extra matter interaction. Such
solution represents an empty spatially homogeneous and isotropic
expanding universe \cite{outros}.

\section*{Acknowledgements}
The major part of this work was done when visiting the International
Center for Relativistic Astrophysics (ICRANet) at Pescara (Italy). I
warmly thank the scientific staff with whom I discussed during
October 2006 and particularly Remo Ruffini. I would extend this
acknowledgement to its efficient administration. I would also thank
my colleagues of ICRA-Brasil and particularly Dr J.M. Salim with
whom I exchanged many discussions concerning the Spinor Theory of
Gravity.  I would like to thank Dr Samuel Senti  for his kind help
in the final English version of this manuscript.

This work was partially supported by {\em Conselho Nacional de
Desenvolvimento Cient\'{\i}fico e Tecnol\'ogico} (CNPq) and {\em
Funda\c{c}\~ao de Amparo \`a Pesquisa do Estado de Rio de Janeiro}
(FAPERJ) of Brazil.


\begin{thebibliography}{100}
\bibitem{will2006} C. M. Will in Living Rev. Relativity, 9, (2006), 3.
\bibitem{stg1} M. Novello, Spinor Theory of Gravity. ArXiv
gr-qc/0609033.
\bibitem{detf} R Kolb et al. DRAFT/ Dark Energy Task Force
(preprint) March, 2006.
\bibitem{Elbaz} E. Elbaz, The Quantum Theory of Particles, Fields and
Cosmology. Editor Springer, 1998. See also V. W. Kofink, Ann. der
Phys, Band 30, 91 (1937).
\bibitem{comentario} In the previous work I used the restricted
definition $X \equiv J^{\mu} J_{\mu} + j^{\mu} j_{\mu}.$ I decided
to change for the present definition in order to include in the same
expression the case in which the field $\Psi$ is an eigen state of
$\gamma^{5}.$
\bibitem{sabado} We note that when Minkowski metric is written in an
arbitrary system of coordinates it is represented by
$\gamma_{\mu\nu}$.
\bibitem{landau} L. Landau and E. Lifshitz, Theory of field
\bibitem{jauch} This remark of Pauli was commented
to me in a private conversation with professor J.M.Jauch in the
seventies.
\bibitem{pauli1} Actually the argument was given in the system of units in which
$\hbar = c = 1.$
\bibitem{Feynman} Feynman Lectures on Gravitation, R. P. Feynman
(Addison-Wesley Publ. Co., 1995).
\bibitem{GPP} L. P. Grishchuk, Sov. Phys. Usp. 33 (8) 669, August
1990. See also LP Grishchuk, A N Petrov and A A Popova in Commun.
Math. Phys. 94 (1984) 379.
\bibitem{GPPdesitter} In \cite{GPP} another choice is investigated, concerning the
deSitter geometry.
\bibitem{heisenberg1} W.Heisenberg
: Rev Mod Phys vol 29 (1957) 269.
\bibitem{outros}  After this work was completed, some papers
concerning related matters were pointed out to me. Among these we
can quote the fusion theory of de Broglie; the non linear spinor
theory of Stumpf et al.; the Rishon model of Harari et al and the
interesting idea of composite graviton by C. Heuson. Let us stress
that, although the STG shares some similarities with the theories
mentioned above, they differ in some crucial aspects:
 these theories aims to give dynamical equations for the
 gravitational field. Nothing similar is done in the
 present Spinor Theory of Gravity. The gravitational field in STG is just the way
 fields  $\Psi$ and $\Upsilon$ couple with matter and
 consequently there is no such thing "a dynamical equation for the
 gravitational field".
 Instead, it comes as a byproduct of the dynamical equation of the
 fundamental spinors. Gravity does not have a dynamics of its own.
 Gravity is just an effective theory.
\end{thebibliography}
 \end{document}